\documentclass[preprint,aps]{revtex4}
\usepackage {epsfig,color}
\usepackage{graphicx}
\usepackage{bm}

\begin{document}


\title{Magnetic and phonon mechanisms of superconductivity in La$_{2-x}$Sr$_x$CuO$_4$ support each other.}

\author {S.G.~Ovchinnikov}
 \email {sgo@iph.krasn.ru}
\author {E.I.~Shneyder}

 \affiliation {L.V. Kirensky Institute of Physics, Siberian Branch of Russian Academy of Sciences, 660036 Krasnoyarsk, Russia}

\date{\today}

\begin{abstract}
Strong electron correlations are responsible both for the insulator ground state of undoped La$_2$CuO$_4$ and strong antiferromagnetic coupling $J$  between neighbouring spins. We consider magnetic mechanism of superconducting pairing in the effective low energy $t - t' - t'' - J^*$ model with all parameters calculated {\it ab initio}. Interaction of strongly correlated electrons with different phonon modes is also incorporated. In a BCS type theory the $d_{x^2  - y^2 }$ gap is given by a sum of magnetic and phonon contributions. The phonon coupling parameter $\lambda  = f(x)G$, where $G$ is a combination of bare electron-phonon couplings for all modes and the function $f$ depends on the hole concentration $x$ due to strong electron correlations. The main contribution to the only fitting parameter $G$ is determined by a competition of the breathing and buckling modes. Fitting the parameter $G$ from the isotope effect we obtain that magnetic and phonon contributions to the critical temperature $T_c $ work together and are of the same order of magnitude.
\end{abstract}


\maketitle

\section{\label{sec:level1}Introduction}

The discovery of superconductivity in cuprates with high critical temperature $T_c $ and  $d$-wave order parameter has serious implications for the theory of superconductivity. The presence of large Coulomb interactions that have the potential to destroy conventional  $s$-wave BCS state and to transform the half-filled single electron band in undoped La$_2$CuO$_4$ into the Mott-Hubbard insulator has prompted the search for new approaches for the normal and superconducting phases. A microscopic approach based on the conventional {\it ab initio} local density approximation (LDA) is not valid in the underdoped region of the cuprate phase diagram due to the strong electron correlations (SEC) effects~\cite{WangPickett0}. In general the problem of strong electron correlations is not solved. There are many different theoretical approaches to this problem including Quantum Monte Carlo, Exact Diagonalization for finite cluster, Slave Boson, Cell Perturbation Theory, Dynamical Mean Field theory for infinite lattice~\cite{Maekawa1,Dagotto1,Tremblay1,Anderson1,Kotliar1,Izyumov1,Tokura1}.

The multielectron generalized tight-binding (GTB) method has been proposed~\cite{Ovchinnikov2} which describes electrons as the quasiparticle excitations between the local multielectron configurations with the interatomic hopping resulting in the dispersion and the band structure. The GTB approach has been successful for undoped cuprates band structure~\cite{Gav3}. The hybrid LDA+GTB method~\cite{Kor4} takes the advantages of both {\it ab initio} single electron and multielectron approaches. The low energy effective Hamiltonian generated by the quasiparticle electronic structure is given by the $t - t' - t'' - J^*$ model (here star means that the three-cite correlated hopping $\sim J$ is included) with all parameters calculated {\it ab initio}~\cite{Kor4}. The normal state of the underdoped cuprate is characterized by electron hopping in the spin liquid background. At small doping in the antiferromagnetic phase hole is a spin polaron and at larger doping spin fluctuations of the short magnetic order also strongly renormalized hole dispersion~\cite{Fulde5,Eder6,Schreiber7}. A spin fluctuation pseudogap has been found both in the spin-fermion model~\cite{Pines8}, in the doped Mott-Hubbard insulator by the cellular dynamical mean field theory (DMFT)~\cite{Kyung9}, and in recent LDA+DMFT+$\Sigma _{\bf{k}}$ calculations~\cite{Sadovskii10} for Bi2212.

A self-consistent consideration of the electronic structure and spin correlation functions within the $t - t' - t'' - J^*$ model with parameters of La$_{2-x}$Sr$_x$CuO$_4$ results in the doping evolution of the hole Fermi surface from small pockets around $\left( {{\pi  \mathord{\left/
 {\vphantom {\pi  2}} \right.
 \kern-\nulldelimiterspace} 2}{{,\pi } \mathord{\left/
 {\vphantom {{,\pi } 2}} \right.
 \kern-\nulldelimiterspace} 2}} \right)$ at $x < x_{cr} $ to large hole surface around $\left( {\pi ,\pi } \right)$ at $x > x_{cr} $ with quantum phase transition~\cite{Kor11} at $x_{cr}  \approx 0.15$. In this paper we extent this approach to the superconducting phase taking into account both magnetic pairing within the $t - t' - t'' - J^*$ model and electron-phonon interaction (EPI). In spite of large number of different phonon modes only a few of them have large EPI including the apical oxygen breathing mode (apical oxygen displacement perpendicular the CuO$_2 $ plane which modifies the Madelung energy), the in-plane oxygen breathing/half-breathing mode and the buckling mode with in-plane oxygen ions moving perpendicular to the   plane~\cite{Bulut12,Nunner13}. Recent {\it ab initio} study of the electronic structure and EPI in La$_{2-x}$Sr$_x$CuO$_4$ has proved that these three modes contribution to the hole self-energy is more than ${\rm{80\% }}$ of the total self-energy~\cite{Giustino14}.
 
The multiband $pd$-model at low energies is reduced to the effective Hubbard model with two Hubbard subbands. The lower and upper Hubbard bands (LHB and UHB) are the bands of the Hubbard fermions created by the $X$-operators $X_f^{0\sigma } $ and $X_f^{ - \sigma ,2} $, correspondingly, $X_f^{pq}  = \left| p \right\rangle \left\langle q \right|$. Here $\left| 0 \right\rangle ,\;{\rm{ }}\left| {\sigma  =  \pm 1/2} \right\rangle $ and $\left| 2 \right\rangle $ are the multielectron eigenstates of the $CuO_6$ unit cell, corresponding to configurations $ d^{10} p^6 $ with number of holes $n_h  = 0$, $d^9 p^6  + d^{10} p^5 $ with $n_h  = 1$, and $d^8 p^6  + d^9 p^5  + d^{10} p^4 $ with $n_h  = 2$. In the hole language the electrons in the valence band correspond to the holes in the UHB. The amplitudes of interatomic hopping in the LHB and UHB are $t_{fg}^{00}$ and $t_{fg}^{11} $, while the interband hopping is given by $t_{fg}^{01} $. When we eliminate the interband excitation throw the charge transfer gap $U_{eff} $ by the standard unitary transformation the effective Hamiltonian for holes in the UHB is given by~\cite{Bulaevskii,ChaoSpalek} $H_{t - J^*}  = H_{t - J}  + H_{(3)}$ with
\begin{eqnarray*}
H_{t - J}  = \sum\limits_{f\sigma } {(\varepsilon  - \mu )} X_f^{\sigma \sigma }  + \sum\limits_{fg\sigma } {t_{fg}^{11} } X_f^{2, - \sigma } X_g^{ - \sigma ,2}  +  \nonumber\\
\sum\limits_{fg} {J_{fg} \left( {\vec S_f  \cdot \vec S_g  - {\textstyle{1 \over 4}}n_f n_g } \right)}, \nonumber\\
H_{(3)}  = \sum\limits_{fmg\sigma } {\frac{{t_{fm}^{01} t_{mg}^{01} }}{{U_{eff} }}\left( {X_f^{2,\bar \sigma } X_m^{\sigma ,\sigma } X_g^{\bar \sigma ,2}  - X_f^{2,\sigma } X_m^{\sigma ,\bar \sigma } X_g^{\bar \sigma ,2} } \right)} . 
\end{eqnarray*}
Where $ J = {t_{01}^{2} } /U_{eff} $ is the super exchange interaction. Distance dependent hopping parameters $t_{fg}^{11}$ have been calculated up to 6-th neighbours and it was revealed that contributions of the fourth and more distant neighbours are neigligible small~\cite{Kor4}. This is the microscopic justification of the $t - t' - t'' - J^*$ model with $3$ hopping parameter (${\rm{ t  = 0}}{\rm{.932}}$, ${\rm{ t'  =  - 0}}{\rm{.120}}$, ${\rm{t'' = 0}}{\rm{.152}}$, ${\rm{J  = 0}}{\rm{.298}}$, ${\rm{J'  = 0}}{\rm{.003}}$, ${\rm{ J''  = 0}}{\rm{.007}}$), all parameters are in eV). The last term $H_{\left( 3 \right)}$ corresponds to the three-site correlated hopping that has the same order as the exchange term $J$ and has to be included in the theory of superconductivity~\cite{Valkov23}.

In the strong electron correlation regime the EPI is the interaction of phonon (with wave number ${\bf{q}}$, frequency $\omega _{{\bf{q}}\nu } $, and mode $\nu$) and Hubbard fermions~\cite{OvchShneyd15}. The effective total Hamiltonian is given by
\begin{equation}
H_{eff}  = H_{t - J^*}  + H_{el - ph - el} 
\end{equation}
where effective electron-electron interaction mediated by phonons is $H_{el - ph - el}  = \sum\limits_{{\bf{kk'q}}} {\sum\limits_{\sigma \sigma '} {V_{{\bf{kk'q}}} X_{{\bf{k}} + {\bf{q}}}^{2,\bar \sigma } X_{{\bf{k}}' - {\bf{q}}}^{2,\bar \sigma '} X_{{\bf{k}}'}^{\bar \sigma ',2} X_{\bf{k}}^{\bar \sigma ,2} } } $ with the effective interaction neglecting retardation effects given by $V_{{\bf{kk'q}}}  =  - \sum\limits_v {{{g_v \left( {{\bf{k}}{\bf{,q}}} \right)g_v \left( {{\bf{k}}'{\bf{,}} - {\bf{q}}} \right)} \mathord{\left/ {\vphantom {{g_v \left( {{\bf{k}}{\bf{,q}}} \right)g_v \left( {{\bf{k}}'{\bf{,}} - {\bf{q}}} \right)} {\omega _{{\bf{q}},v} }}} \right. \kern-\nulldelimiterspace} {\omega _{{\bf{q}},v} }}} $.

For isotropic $s$-wave gap all phonon modes contributed additively to pairing. For anisotropic $d_{x^2  - y^2 } $-gap the wave number dependence of the EPI matrix elements is crucial. Maximal EPI for the breathing/half-breathing mode at large ${\bf{q}} \sim {\pi  \mathord{\left/ {\vphantom {\pi  a}} \right. \kern-\nulldelimiterspace} a}$ results in depairing effect of this mode, while the buckling mode with maximum of interaction at ${\bf{q}} = 0$  supports the $d_{x^2  - y^2 } $ pairing. This conclusion has been obtained by different approaches~\cite{Bulut12,Nunner13,Shneyder16,Honerkamp17} and results from a simple physics: large ${\bf{q}}$ EPI changes the phase of the $d_{x^2  - y^2 } $-gap on the Fermi surface while small ${\bf{q}}$ EPI does not change the phase.

It should be emphasized that mean-field theory in the framework of GTB and GTB+LDA methods differs from standard mean field treatment of the Coulomb interaction like $Un_1 n_2  \to Un_1 \left\langle {n_2 } \right\rangle $, where $n_i $ is a number of particles in $i$ state. In these methods the cluster perturbation theory is used that combines the exact diagonalization treatment of the multiband $pd$-model Hamiltonian inside the unit cell, and perturbation treatment of the intercell hopping in the $X$-operator representation. For the normal state this mean field approach is just a cluster generalization of the Hartree-Fock approximation for Hubbard fermions. For superconducting state, the mean field theory has been developed in the $X$-operator representation that is reliable in the strong correlation regime~\cite{Plakida18}. Double occupation is prohibited in this approach by the local constraint formulated in the $X$-operator representation similar to the local constraint in the slave boson approach. Contrary to the slave boson mean field theory where the local constraint is violated, in our mean field theory the $X$-operator algebra provides the local constraint in all stages of calculations.

\section{isotope effect}
In the BCS-type approach to superconductivity a spin singlet pairing of the Hubbard fermions is given by the anomalous average~\cite{Plakida18} $B_{\bf{q}}  = \left\langle {X_{ - {\bf{q}}}^{\sigma ,2} X_{\bf{q}}^{\bar \sigma ,2} } \right\rangle $. For the $d_{x^2  - y^2 } $-pairing the gap equation reads~\cite{Shneyder16}
\begin{eqnarray}
\label{eq_gap}
\begin{array}{l}
 \Delta _{\bf{k}}  = \frac{{2\varphi _{\bf{k}} }}{N}\sum\limits_{\bf{q}} {\left\{ {\frac{{1 - x}}{2}J + \lambda \theta \left( {\left| {\xi _{\bf{q}}  - \mu } \right| - \omega _D } \right)} \right\}}  \frac{{2\Delta _{\bf{q}} \varphi _{\bf{q}} }}{{\xi _{\bf{q}}  - \mu }}\tanh \left( {\frac{{\xi _{\bf{q}}  - \mu }}{{2\tau }}} \right) \\ 
 \end{array} 
\end{eqnarray}
where $\tau  = k_B T$, $k_B $ is Boltzmann constant, and $T$ is temperature, $\varphi _{\bf{q}}  = {{(\cos q_x a - \cos q_y a)} \mathord{\left/ {\vphantom {{(\cos q_x a - \cos q_y a)} 2}} \right. \kern-\nulldelimiterspace} 2}$ is the angle-dependent part of the gap $\Delta _{\bf{q}}  = \Delta _0 \varphi _{\bf{q}} $, the total coupling parameter in brackets is given by a sum of magnetic $J$ and phonon $\lambda $ couplings. The $\theta $-function as usually means that phonon pairing occurs in a narrow energy window of the $\omega _D $ width near the Fermi energy. The normal phase dispersion $\xi _{\bf{q}} $ takes into account the spin correlation function $c_{\bf{q}}$, and three-cite interaction, the chemical potential $\mu$ is calculated self-consistently for the carrier concentration $x$ in La$_{2-x}$Sr$_x$CuO$_4$. The phonon coupling parameter $\lambda  = f\left( x \right) G$, where dimensionless function $f\left( x \right)$ is given by $f(x) = {{\left( {1 + x} \right)\left( {3 + x} \right)} \mathord{\left/
 {\vphantom {{\left( {1 + x} \right)\left( {3 + x} \right)} 8}} \right.
 \kern-\nulldelimiterspace} 8} - {{3c_{01} } \mathord{\left/
 {\vphantom {{3c_{01} } 4}} \right.
 \kern-\nulldelimiterspace} 4}$, and the parameter $G$ is determined by the bare EPI matrix elements $G = \left( {{{g_{buck}^2 } \mathord{\left/ {\vphantom {{g_{buck}^2 } {\omega _{buck} }}} \right. \kern-\nulldelimiterspace} {\omega _{buck} }} - {{g_{breath}^2 } \mathord{\left/ {\vphantom {{g_{breath}^2 } {\omega _{breath} }}} \right. \kern-\nulldelimiterspace} {\omega _{breath} }}} \right)$.

The concentration dependence of both magnetic and phonon couplings stems from the unusual statistics of the Hubbard fermions. Contrary to the free electron band with two electrons per atom the Hubbard subbands have the odd number of states which depends on concentration via total number of holes $n_h  = 1 + x$ and the nearest neighbour spin correlation function $c_{01} $. Due to the antiferromagnetic type of correlation $c_{01} $ is negative and its contribution to $\lambda $ is positive. The appearance of the spin correlation function in the phonon coupling means some interference of the magnetic and phonon mechanisms of pairing. This function $ c_{01}  = 2\left\langle {S_0^z S_1^z } \right\rangle  = \left\langle {S_0^ +  S_1^ -  } \right\rangle $ characterizes the spin liquid properties of the underdoped cuprate and is concentration dependent~\cite{Kor11}.

All parameters in the gap equation (\ref{eq_gap}) but $G$ have been obtained within {\it ab initio} LDA+GTB approach. The precision of modern calculations of the EPI matrix elements especially for strongly correlated electrons seems to be not enough. Thus, different approaches give opposite conclusions. The larger EPI of the half-breathing mode versus buckling one has been discussed in papers~\cite{Nunner13,Khaliullin19,Rosch20}. The average over Brillouin zone value $\lambda $ for buckling mode was obtained much larger than for breathing/half-breathing one~\cite{Devereaux21}. The {\it ab initio} calculations of the Eliashberg function $\alpha ^2 F$ for YBa$_2$Cu$_3$O$_7$ have revealed rather small EPI parameter $\lambda  = 0.27$ in disagreement with earlier approximate treatments~\cite{Bohnen22}. Thus in this paper we consider the parameter $G$ as the only fitting parameter, to find the value of $G$ we calculate the isotope effect exponent determining as $\alpha _O  =  - \frac{{{\rm{d}}\ln \left( {T_c } \right)}}{{{\rm{d}}\ln \left( {M_O } \right)}} $. Using the equation (\ref{eq_gap}) we get the equation for the superconducting transition temperature $T_c$ then the analytical expression for $\alpha _O $ can be written in the form $\alpha _O  = \frac{{Int1}}{{Int2}}$ where
\begin{eqnarray}
\label{eq_alpha}
\begin{array}{l}
 Int1 = \frac{{\omega _D }}{N}\sum\limits_{\bf{q}} {\frac{{4\varphi _{\bf{q}} ^2 }}{{\xi _{\bf{q}} }}\lambda _{ph} \delta \left( {\left| {\xi _{\bf{q}}  - \mu } \right| - \omega _D } \right)\tanh \left( {\frac{{\xi _{\bf{q}}  - \mu }}{{2k_B T_c }}} \right)}  \\ 
 Int2 = \frac{1}{N}\sum\limits_{\bf{q}} {\frac{{4\varphi _{\bf{q}} ^2 }}{{T_C }}\cosh ^{ - 2} \left( {\frac{{\xi _{\bf{q}}  - \mu }}{{2k_B T_c }}} \right) } \left\{ {\frac{{1 - x}}{2}J + \lambda _{ph} \theta \left( {\left| {\xi _{\bf{q}}  - \mu } \right| - \omega _D } \right)} \right\} \\ 
 \end{array}
\end{eqnarray}

\begin{figure}
\center
\includegraphics[width=0.3\linewidth]{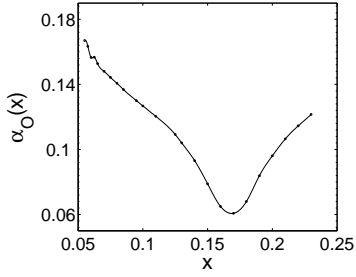}
\caption{\label{fig:alpha} The doping dependence of the oxygen isotope exponent for the effective EPI parameter $G/J=0.35$.}
\end{figure}

\begin{figure}
\includegraphics[width=0.6\linewidth]{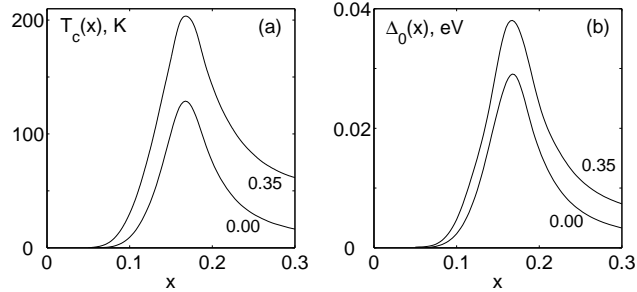}
\caption{\label{fig:TcGap} The critical temperature (a) and the amplitude of the superconducting gap (b) as function of the doping concentration in La$_{2-x}$Sr$_x$CuO$_4$ for pure magnetic $(G=0)$ and both magnetic and phonon mechanisms $(G/J=0.35)$ of pairing.} 
\end{figure}

The calculated oxygen isotope effect exponent $\alpha _O $ as function of hole concentration is shown in the Fig.~\ref{fig:alpha}. We have found that the positive (negative) sign of $G$ results in positive (negative) sign of the exponent $\alpha _O $. The value ${G \mathord{\left/ {\vphantom {G {J = 0.35}}} \right. \kern-\nulldelimiterspace} {J = 0.35}}$ provides $\alpha _O = 0.06$ close to the La$_{2-x}$Sr$_x$CuO$_4$ experimental data at the optimal doping. The increase of the isotope exponent away from the optimal doping is obtained. The higher value of the $G$ parameter would result in the larger $\alpha _O $ above the BCS value $0.5$ but the optimal doping value $\alpha _O \left( {x = 0.17} \right)$ will also increase. The doping dependence of the critical temperature $T_c$ is shown in the Fig.~\ref{fig:TcGap}(a). It is clear that this dependence reproduces well the general structure of the superconducting dome with the optimal doping at $x=0.17$ and disappearance of superconductivity in the underdoped region below  $x=0.06$. In the overdoped region the decrease of $T_c$ is more smooth than in experiment. Our approach from the undoped regime becomes less accurate in the overdoped region then at small doping. We have obtained that with positive parameter $G$ the phonon contribution to pairing increases the magnetic one. The phonon contribution to the critical temperature is a little bit less than magnetic one but of the same order of magnitude. The absolute value of $T_c$ is too large and this is the general drawback of the mean field theory. The gap amplitude $\Delta _0$ as function of the hole concentration is shown in the Fig.~\ref{fig:TcGap}(b). Similar to the Fig.~\ref{fig:TcGap}(a) the phonon contribution increases the gap value. At optimal doping the ratio ${{2\Delta _0 } \mathord{\left/
 {\vphantom {{2\Delta _0 } {k_B T_c }}} \right.
 \kern-\nulldelimiterspace} {k_B T_c }} = 4.9$ is close to the experimental data~\cite{Wang24}.

We would like to discuss the effect of the strong EPI with apical oxygen breathing mode. This interaction is poor screened. The direct experimental proof of its importance is demonstrated by the colossal lattice expansion along  $c$-axis in La$_2$CuO$_{4 + \delta }$ under femtosecond intensive light pulses~\cite{Gedik25}. Nevertheless this large EPI does not contribute to the superconducting $d$-pairing due to the orthogonality of the in-plane electron momentum and  $c$-axis phonon wave number for the apical oxygen breathing mode~\cite{Shneyder16}. The site selective isotope substitution confirms the absence of the isotope effect when the isotope is in the apical oxygen position~\cite{Khasanov27,Bishop28}. Previously the absence of the apical oxygen breathing mode contribution to the $d_{x^2  - y^2 }$ pairing has been  obtained in the paper~\cite{Bulut12}.

The effect of strong electron correlations on the phonon-mediated superconductivity and oxygen isotope effect has been considered in the slave-boson approach to the Hubbard model~\cite{Kim29}. The total agreement of our and slave-boson approaches is not possible. Thus the unphysical result of the zero bandwidth and zero EPI in the undoped limit~\cite{Kim29} is absent in our GTB method which provides the dispersion of the quasiparticles at the top of the valence band in undoped cuprates in agreement with the ARPES experiments. Nevertheless close to the optimal doping we can compare both approaches. Our EPI parameter increases with hole concentration  $x$, and the $\lambda _{pd}$ in the paper~\cite{Kim29}  also increases with $x$. The Coulomb interaction $ \mu _{SB}^* $ decreases~\cite{Kim29}  with  $x$, and our magnetic coupling which is the effect of strong correlations also decreases with $x$. The isotope effect value has similar doping dependences in the paper~\cite{Kim29}  and in our work. The phonon-mediated  $s$-wave superconductivity with similar doping dependent isotope effect have been also discussed in the papers~\cite{Nazarenko30,Chen31}.

The proper doping dependence of the oxygen isotope exponent $ \alpha $ has been obtained recently in the anharmonic version of the buckling mode mediated $ d_{x^2  - y^2 } $ superconductivity~\cite{Newns32}. However this model neglects the depairing effect of the breathing and half-breathing mode and neglects the strong electron correlations which result in the magnetic mechanism of pairing.

Summarizing our discussion we want to emphasize that both magnetic and phonon mechanisms of  $d_{x^2  - y^2 } $-pairing should be considered in realistic theory of superconductivity in cuprates. Many authors have previously discussed separately the magnetic mechanisms of pairing generated by strong electron correlations or phonon pairing which explains the isotope effect. Here we have shown that both mechanisms may work together increasing each other. Our theory is almost parameters-free. It is based on the correct microscopic description of the undoped insulator state of La$_2$CuO$_4$ and the doping evolution of the emerging hole Fermi surface. All parameters of the electronic structure and the magnetic mechanisms of pairing have been calculated within the {\it ab initio} LDA+GTB approach. The only parameter entering our theory is the combination of bare electron-phonon matrix elements $G$. Its sign $G > 0$ is required to have the positive oxygen isotope exponent $\alpha$, its value can be fitted to get the proper concentration dependence of $\alpha \left( x \right)$. The {\it ab initio} calculation of the electron-phonon matrix elements in the regime of strong electron correlations still remains the important unsolved problem.

\begin{acknowledgments}
This work is supported by the Presidium RAS program "Quantum macrophysics", the integration project SORAN-UrORAN \verb+#+74, Project SORAN 3.4, the RFFI Grants 06-02-16100, 06-02-90537-BNTS.
\end{acknowledgments}


\begin{thebibliography}{99}
%

\bibitem{WangPickett0} T.C. Leung, X.W. Wang, and B.N. Harmon, Phys. Rev. B {\bf 37}, 384 (1988); W.E. Pickett, Rev. Mod. Phys. {\bf 61}, 433 (1989).
\bibitem{Maekawa1} S. Maekawa, T. Tohyama, S.E. Barnes, S. Ishihara, W. Koshibae, G. Khaliullin, in {\it Physics of Transition Metal Oxides}, edited by M. Cardona, et al, H. St\"{o}rmer (Springer, New York, 2004).
\bibitem{Dagotto1} E. Dagotto, Rev. Mod. Phys. {\bf 66}, 763 (1994).
\bibitem{Tremblay1} D. S\'{e}n\'{e}chal, A.-M. Tremblay and C. Bourbonnais, in {\it Theoretical Methods for strongly correlated electrons}, CRM Series in Mathematical Physics, (Springer, New York, 2003).
\bibitem{Anderson1} P.W. Anderson, in {\it The theory of Superconductivity in the high-Tc Cuprates} (Princeton University Press, Princeton NJ 1997).
\bibitem{Kotliar1} G. Kotliar, S.Y. Savrasov, K. Haule, V.S. Oudovenko, O. Parcollet, C.A. Marianetti, Rev. Mod. Phys. {\bf 78}, 865 (2006).
\bibitem{Izyumov1} Y.A. Izyumov, E.Z. Kurmaev, Phys. Usp. {\bf 51}, 23 (2008).
\bibitem{Tokura1} M. Imada, A. Fujimori, Y. Tokura, Rev. Mod. Phys. {\bf 70}, 1039 (1998).
\bibitem{Ovchinnikov2} S.G. Ovchinnikov, and I.S. Sandalov, Physica C {\bf 161}, 607 (1989).
\bibitem{Gav3} V.A. Gavrichkov, S.G. Ovchinnikov, A.A. Borisov, and E.G. Goryachev, JETP {\bf 91}, 369 (2000).
\bibitem{Kor4} M.M. Korshunov, V.A. Gavrichkov, S.G. Ovchinnikov, I.A. Nekrasov, Z.V. Pchelkina, and V.I. Anisimov, Phys. Rev. B {\bf 72}, 165104 (2005).
\bibitem{Fulde5} P. Unger, and P. Fulde, Phys. Rev. B {\bf 47}, 8947 (1993).
\bibitem{Eder6} R. Eder, and Y. Ohta, Phys. Rev. B {\bf 50}, 10043 (1994).
\bibitem{Schreiber7} A. Sherman, and M. Schreiber, Eur. Phys. J. B {\bf 32}, 203 (2003).
\bibitem{Pines8} J. Schmalian, D. Pines, and B. Stojkovic, Phys. Rev. B {\bf 60}, 667 (1999).
\bibitem{Kyung9} B. Kyung, S.S. Kancharla, D. S\'{e}n\'{e}chal, A.-M.S. Tremblay, M. Civelli and G. Kotliar, Phys. Rev. B {\bf 73}, 165114 (2006).
\bibitem{Sadovskii10} E.Z. Kuchinskii, I.A. Nekrasov, Z.V. Pchelkina, and M.V. Sadovskii, JETP {\bf 10}4, 792 (2007).
\bibitem{Kor11} M.M. Korshunov, and S.G. Ovchinnikov, Europ. J. Physics  {\bf 57}, 271 (2007).
\bibitem{Bulut12} N. Bulut, and D.J. Scalapino, Phys. Rev. B {\bf 54}, 14971 (1996).
\bibitem{Nunner13} T.S. Nunner, J. Schmalian, and K. H. Bennemann, Phys. Rev. B {\bf 59}, 8859 (1999).
\bibitem{Giustino14} F. Giustino, M.L. Cohen, and S.G. Louie, Nature {\bf 452}, 975 (2008).
\bibitem{Bulaevskii} L. Bulaevskii, E. Nagaev, and D. Khomskii, JETP {\bf 27}, 836 (1968). 
\bibitem{ChaoSpalek} K.A. Chao, J. Spalek, and A.M. Oles, J. Phys. C {\bf 10}, L271 (1977).
\bibitem{Valkov23} V.V. Val'kov, T.A. Val'kova, D.M. Dzebisashvili, and S.G. Ovchinnikov, JETP Lett. {\bf 75}, 378 (2002).
\bibitem{OvchShneyd15}  S.G. Ovchinnikov, and E.I. Shneider, JETP {\bf 101}, 844 (2005).
\bibitem{Shneyder16} E.I. Shneyder, and S.G. Ovchinnikov, JETP Lett.{\bf 83}, 394 (2006).
\bibitem{Honerkamp17} C. Honerkamp, H.C. Fu, and D.-H. Lee, Phys. Rev. B {\bf 75}, 014503 (2007).
\bibitem{Plakida18} N.M. Plakida, and V.S. Oudovenko, Phys. Rev. B {\bf 59}, 11949 (1999).
\bibitem{Khaliullin19} G. Khaliullin, and P. Horsch, Phys. Rev. B {\bf 54}, R9600 (1996).
\bibitem{Rosch20} O. Rosch, and O. Gunnarsson, Phys. Rev. Lett. {\bf 92}, 146403 (2004).
\bibitem{Devereaux21} T.P. Devereaux, T. Cuk, Z.-X. Shen, and N. Nagaosa Phys. Rev. Lett. {\bf 93}, 117004 (2004).
\bibitem{Bohnen22} K.-P. Bohnen, R. Heid, and M. Krauss, Europhys. Lett. {\bf 64}, 104 (2003).
\bibitem{Wang24} Y. Wang, J. Yan, L. Shan, H.-H. Wen, Y. Tanabe, T. Adachi, and Y. Koike, Phys. Rev. B {\bf 76}, 064512 (2007).
\bibitem{Gedik25} N. Gedik, D.-S. Yang, G. Logvenov, I. Bozovic, and A. H. Zewail, Science {\bf 316}, 425 (2007).
\bibitem{Khasanov27} R. Khasanov, et al, Phys. Rev. B {\bf 68}, 220506(R) (2003).
\bibitem{Bishop28} A.R. Bishop, et al, J. Supercond. Nov. Magn. {\bf 20}, 393 (2007).
\bibitem{Kim29} J.H. Kim, and Z. Tesanovic, Phys. Rev. Lett. {\bf 71}, 4218 (1993).
\bibitem{Nazarenko30} A. Nazarenko, and E. Dagotto, Phys. Rev. B {\bf 53}, R2987 (1996). 
\bibitem{Chen31} X.-J. Chen, V.V. Struzhkin, Z. Wu, R. J. Hemley, and H.-k. Mao, H.-Q. Lin, Phys. Rev. B {\bf 75}, 134504 (2007).
\bibitem{Newns32} D.M. Newns, and C.C. Tsuei, Nature Physics {\bf 3}, 184 (2007).
%
\end{thebibliography}
\end{document}